\documentclass[
pra,aps,
reprint,
a4paper,
noeprint,
amsmath,amssymb,amsbsy,
superscriptaddress,
floatfix
]{revtex4-2}
\usepackage{graphicx}
\usepackage{dcolumn}
\usepackage{bm}

\usepackage{etoolbox}
	\newtoggle{arXiv} \toggletrue{arXiv}
	\newtoggle{bibrun}

\usepackage{siunitx, physics}
	\newcommand{\I}{\ensuremath{\mathrm{i}}}
	\newcommand{\Exp}[1]{\mathrm{e}^{#1}}

\begin{document}


\title{Generation of a maximally entangled state using collective optical pumping}

\author{M. Malinowski}
\email{maciejm@phys.ethz.ch}
\author{C. Zhang}
\author{V. Negnevitsky}
\author{I.~Rojkov}
\author{F. Reiter}
\author{T.-L.~Nguyen}
\author{M. Stadler}
\author{D. Kienzler}
\author{K. K. Mehta}
\affiliation{Institute for Quantum Electronics, ETH Z\"urich, 8093 Z\"urich, Switzerland}
\author{J. P. Home}
\email{jhome@phys.ethz.ch}
\affiliation{Institute for Quantum Electronics, ETH Z\"urich, 8093 Z\"urich, Switzerland}
\affiliation{Quantum center, ETH Z\"urich, 8093 Z\"urich, Switzerland}


\begin{abstract}
We propose and implement a novel scheme for dissipatively pumping two qubits into a singlet Bell state. The method relies on a process of collective optical pumping to an excited level, to which all states apart from the singlet are coupled. We apply the method to deterministically entangle two trapped ${}^{40}\text{Ca}^+$ ions with a fidelity of $93(1)\%$. We theoretically analyze the performance and error susceptibility of the scheme and find it to be insensitive to a large class of experimentally relevant noise sources. 
\end{abstract}

\maketitle


%
Quantum entanglement is a resource for quantum computation \cite{Jozsa2003}, communication \cite{Bennett1992}, cryptography \cite{Ekert1991} and metrology \cite{Toth2014}. Entangled states of bipartite systems are typically prepared using a two-step process, the first involving initialization of a separable state by optical pumping, followed by a unitary transformation which generates entanglement \cite{DiVincenzo2000}. In such an \emph{open-loop process} the final state is sensitive to the pulse parameters used to create it and is not protected from future errors. An alternative mode of operation is to use a \emph{closed-loop process}, where feedback from a low-entropy reference system drives the system continuously towards the desired state or subspace. This can be done using measurement-conditioned classical control (e.g. quantum error correction or outcome heralding) or through dissipative engineering, for which the reference is provided by a zero-occupation reservoir \cite{Poyatos1996, Kraus2008, Plenio1999, Verstraete2009, Ticozzi2014}. Dissipation engineering allows useful quantum states to be created in the steady-state, making the process self-correcting with regard to transient errors \cite{Kastoryano2011, Morigi2015}, and resulting in a resource state or subspace which is continuously available. Entanglement of qubits using dissipative engineering has previously been demonstrated using trapped ions \cite{Barreiro2011, Lin2013}, atomic ensembles \cite{Krauter2011}, and superconducting circuits \cite{Shankar2013, Liu2016, Kimchi-Schwartz2016}.
Beyond qubit-based approaches, reservoir engineering has been used to create and stabilize non-classical states of bosonic systems \cite{Poyatos1996, Kienzler2017, DeNeeve2020} as well as to perform quantum error-correction \cite{DeNeeve2020, Gertler2021}.

A widely used strategy for dissipation engineering is to rely on engineered resonances, whereby pumping into the desired entangled state is achieved by resonant drives, while leakage processes out of the desired state are off-resonant \cite{Vacanti2009, Kastoryano2011, Reiter2012, Lin2013}. This approach has proven to be versatile, and has been theoretically extended to the generation of multi-qubit states \cite{Cho2011, Lin2016, Reiter2016}, quantum error correction \cite{Cohen2014, Reiter2017}, and quantum simulation \cite{Reiter2021}. However, dissipative protocols based on resonance engineering can be slow to converge. This is due to the fact that, in order to suppress leakage processes, the drives need to be weak compared to the splittings of the resonances. The resulting competition with additional uncontrolled dissipation channels thus limits the achievable fidelities. It has been proposed that this issue could be overcome by dissipative schemes based on symmetry \cite{Bentley2014, Horn2018, Doucet2020, Cole2021}.

\begin{figure}[b]
\includegraphics{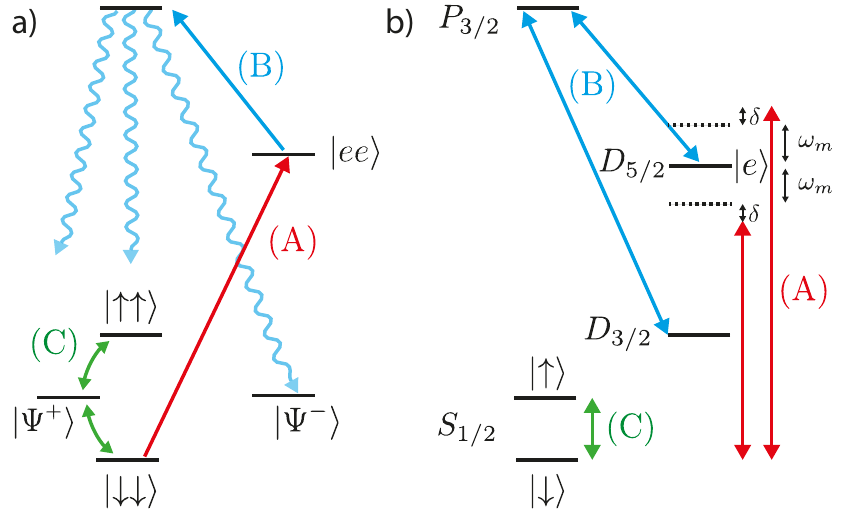}
\caption{(a) High-level description of the protocol. When drives (A), (B), and (C) are switched on, the system is pumped into a maximally entangled state $\ket{\Psi^-}$. (b) Atomic transitions and drives in ${}^{40}\text{Ca}^+$ used in the implementation described in this work. (A) and (B) are driven by laser beams, while (C) is implemented by a global oscillating B-field. Dashed lines denote motional sidebands of the $\ket{\downarrow} \leftrightarrow \ket{e}$ transition.}
\label{fig:scheme}
\end{figure}

In this Letter, we present a method for dissipatively generating two-body entanglement using a deterministic collective optical pumping process which does not couple to the target entangled state: the singlet Bell state $\ket{\Psi^-} \equiv \left(\ket{\uparrow \downarrow} - \ket{\downarrow \uparrow}\right)/\sqrt{2}$. Unlike previous demonstrations, our method relies on symmetry, involving only global fields which couple equally to each system. We thereby overcome the speed limitations of previous schemes, achieving a faster convergence. Our scheme is robust to global error processes. We implement the protocol using two trapped ions in a surface-electrode trap with integrated optical control fields \cite{Mehta2020}, achieving a $93(1)\%$ fidelity with the desired singlet state. Compared to earlier trapped-ion approaches, our method has the advantage of not requiring ground-state cooling.


The scheme is illustrated using the relevant states of two multi-level systems in Fig~\ref{fig:scheme} a). We consider a spin ground state manifold consisting of the collective spin states $\ket{\downarrow\downarrow}, \ket{\uparrow \uparrow}$, $\ket{\Psi^+} \equiv \left(\ket{\uparrow \downarrow} + \ket{\downarrow \uparrow}\right)/\sqrt{2}$ (spin triplet) and $\ket{\Psi^-}$ (spin singlet), as well as excited states, of which the most important for our purposes consists of both systems in a particular excited state $\ket{e}$. Three elements define the pumping process. 
The first is a collective excitation (A) from the state $\ket{\downarrow \downarrow}$ to the doubly excited state $\ket{ee}$. Its collective nature means that it does not couple to the other states in the ground-state manifold. The state $\ket{ee}$ is quenched through a decay channel (B) that acts independently on the qubits, thus redistributing population from $\ket{ee}$ into all four spin states. The combination of (A) and (B) provides a collective pumping that moves the ground state population from $\ket{\downarrow\downarrow}$ to the other ground states. To prepare only the singlet, this is supplemented by a symmetric drive (C) which resonantly drives both qubits with equal amplitude and phase. Due to its symmetry, this drive cycles population within the triplet subspace, while leaving the singlet untouched. Thus the triplet states have a chance of being repumped through the collective pumping, while population in the singlet is dark to all drives. $\ket{\Psi^-}$ then becomes the steady-state of the system. The protocol can be implemented in a continuous manner or by sequentially applying each component. For our implementation, we expect the latter to be more robust to experimental imperfections and proceed to analyze this case below. The continuous implementation is analyzed in detail in Supp.Mat.~\ref{sec:continuous_scenario}.

To identify optimal settings, we optimize a super-operator which combines the three drives. For the collective excitation this is derived from a unitary 
\begin{align}
\label{eq:MS_unitary_closed}
    U_A(\Phi)= \Exp{-\I \Phi S_{x,e}^2},
\end{align}
with $S_{x,e} = \sigma_{x,\downarrow e}\otimes\bm{1} + \bm{1}\otimes \sigma_{x,\downarrow e}$,  $\sigma_{x,\downarrow e}= \ket{e}\bra{\downarrow} +  \ket{\downarrow}\bra{e}$ and $\bm{1}$ is a $3\times3$ identity operator. This provides a full transfer from $\ket{\downarrow\downarrow}$ to $\ket{ee}$ for $\Phi = \pi/4$. Drive (B) repumps the population from $\ket{e}$ to the qubit states with branching ratios which we parameterize by 
$p_{e\rightarrow \downarrow}/p_{e\rightarrow \uparrow} = \tan^2(\gamma)$. Drive (C) is described by a unitary $U_{C}(\theta) = \exp(\I \frac{\theta}{2}\sigma_x)\otimes \exp(\I \frac{\theta}{2}\sigma_x)$, where
$\sigma_{x} = \ket{\uparrow}\bra{\downarrow} +  \ket{\downarrow}\bra{\uparrow}$. After $N$ cycles of the protocol, the singlet error, defined as $\epsilon = 1-F(\ket{\Psi^{-}})$ with $F(\ket{\Psi^{-}}) = \bra{\Psi^{-}}\rho\ket{\Psi^{-}}$ decays as $\epsilon \propto \exp(-N/N_0)$. Through eigenvalue analysis we find the most rapid convergence for $\Phi = \pi/4$, $\theta\approx0.72\pi$ and $\gamma\approx0.22\pi$ where $N_0=7.62$ cycles (Supp. Mat.~\ref{sec:convergence_rate}). The steady state is insensitive to the values of $\Phi$, $\gamma$ and $\theta$, hence indicating that these parameters do not require precise calibration. 


We implement the protocol on a pair of ${}^{40}\text{Ca}^+$ ions confined in the surface-electrode radio-frequency trap described in \cite{Mehta2020}. The qubit is encoded into ground-state Zeeman sub-levels $\ket{\downarrow} = \ket{S_{1/2}, m_j=-1/2}$ and $\ket{\uparrow} = \ket{S_{1/2}, m_j=+1/2}$ which have a frequency splitting of $2 \pi \times \SI{16.5}{\MHz}$ in the applied magnetic field of \SI{0.59}{\milli\tesla}. We use an ancilliary state $\ket{e} = \ket{D_{5/2}, m_j=-1/2}$. Narrow-linewidth laser light at \SI{729}{\nm} is delivered through trap-integrated photonics, and coherently drives transitions between the $S_{1/2}$ and $D_{5/2}$ levels. Free-space laser beams are used for cooling, repumping, and readout operations. The $\ket{\downarrow}\leftrightarrow\ket{\uparrow}$ transition is driven by resonant radio-frequency magnetic fields.

The collective excitation step (A) is implemented using a bichromatic \SI{729}{\nm} laser field with Rabi frequency $\Omega$ and two frequency components detuned by $\delta = \pm 2 \pi \times \SI{14.7}{\kHz}$ from the red and blue motional sidebands of the $\ket{\downarrow}\leftrightarrow \ket{e}$ transition, using the axial stretch mode at $\omega_m \approx 2\pi \times \SI{2.4}{\MHz}$ for which the Lamb-Dicke parameter $\eta = 0.026$. This results in a Hamiltonian $H_A = \frac{1}{2} \hbar \eta \Omega S_{x,e} (\hat{a} \Exp{\I \delta t} + \hat{a}^{\dag} \Exp{-\I \delta t})$ which implements a force on the oscillator whose phase depends on the eigenstate of $S_{x,e}$. This is commonly referred to as a M{\o}lmer - S{\o}rensen drive, and is one of the primary methods for performing two--qubit gates with trapped ions \cite{Sorensen1999, Benhelm2008a, Gaebler2016a}. A pulse of duration $t$ then results in the unitary
\begin{align}
\label{eq:MS_unitary_general}
U_A = \Exp{(\alpha(t) \hat{a}^\dagger -\alpha^*(t) \hat{a})S_{x,e}} \Exp{\I \Phi(t) S_{x,e}^2},
\end{align}
where $\alpha(t) = -\I \frac{\eta \Omega}{\delta} \Exp{-\I \delta t/2} \sin(\delta t/2)$ is an oscillator phase-space displacement amplitude and $\Phi(t) = \frac{\eta^2 \Omega^2}{4 \delta^2}(\delta t - \sin(\delta t))$ is a collective phase factor. Eq.~\eqref{eq:MS_unitary_general} reduces to a pure $S_{x,e}^2$ coupling of the form of Eq.~\eqref{eq:MS_unitary_closed} in two cases. The first, appropriate to a continuous implementation (Supp. Mat.~\ref{sec:continuous_scenario}), is when $|\delta| \gg \eta \Omega$ and so the oscillator excitation can be neglected \cite{Kim2010}. The second, which is the main focus of this work, is when $t = 2 n \pi/\delta$ with $n \in \mathbb{Z}$, for which $\Phi = n \pi \eta^2 \Omega^2/(2\delta^2)$ \cite{Kirchmair2009c}. Repump (B) is implemented using a laser at \SI{854}{\nm}, which couples all $D_{5/2}$ sub-levels to the short-lived $P_{3/2}$ states, which primarily decay into the ground state manifold. A second decay channel to the $D_{3/2}$ states is repumped using a second laser at \SI{866}{\nm}. After \SI{5}{\micro\second}, we measure a probability of leaving $D_{5/2}$ of $>0.9999$, and no detectable loss outside the $\{S_{1/2},D_{5/2}\}$ subspace. The primary repump channel $\ket{e} \rightarrow \ket{P_{3/2}, m_j=-1/2}$ results in branching ratios of $p_{e\rightarrow \downarrow}\approx 2/3$ and $p_{e\rightarrow \uparrow}\approx 1/3$ ($\gamma \simeq 0.3\pi$). The symmetric drive (C) is implemented by passing a current through a track on a circuit board at around 1~mm distance from the ions, which is well in excess of the \SI{5}{\micro\metre} ion spacing. This results in a negligible difference between the single-ion Rabi frequencies.

The protocol is robust to many global errors but amplifies local error channels (this is discussed later). To mitigate these, a number of coherent control techniques are used. We implement the collective excitation step as a sequence of two pulses ($t = 2 n \pi/\delta$ with $n=2$) with $\eta \Omega = \delta/2$, resulting in $\Phi = \pi/4$ and a drive time of $t=\SI{150}{\micro\second}$. The phase of the force acting on the oscillator is shifted by $\pi$ for the second pulse, thus cancelling any residual displacement produced by a single pulse \cite{Hayes2012}. We were surprised to find that the differential AC Stark shift of the $\ket{\downarrow}\leftrightarrow \ket{\uparrow}$ transition produced by the collective drive (A) is different by $\approx 2 \pi \times \SI{2.5}{\kHz}$ on the two ions, causing a near-complete failure of the protocol (see Supp. Mat. \ref{sec:differential_phase_shift}). To mitigate this, we replace the optimal value of $\theta$ applied in each cycle with two values, $\theta_1 = \pi$ applied in odd cycles (drive time $t_C = \SI{6.4}{\micro\second}$) and $\theta = \pi/2$ ($t_C = \SI{3.2}{\micro\second}$) applied in even cycles. This has the effect of a spin-echo, mitigating the error, but at the cost that high-fidelity singlet states are produced only after even cycles. One cycle of the protocol takes $\approx \SI{165}{\micro\second}$ on average. In the absence of other errors, the protocol produces $\ket{\Psi^-}$ regardless of the ions' temperature. However, finite temperature amplifies existing local errors associated with residual oscillator excitation (i.e. when $|\alpha(t)|>0$) and spectator mode excitation. For this reason, it is practically beneficial to cool the ion close to the motional ground state. 

We measure the $P(\downarrow \downarrow), P(\downarrow \uparrow) + P(\uparrow \downarrow)$, and $P(\uparrow \uparrow)$ populations by shelving $\ket{\downarrow}$ (for both ions) into ancillary $D_{5/2}$ sub-levels, followed by state-dependent fluorescence \cite{Myerson2008a}. This allows us to extract the ground state parity $\langle \sigma_{z} \sigma_{z}\rangle$, while 
$\langle \sigma_{x} \sigma_{x}\rangle$ and $\langle \sigma_{y} \sigma_{y}\rangle$ are obtained by measuring the parity following radio-frequency spin rotations $\exp(\I \frac{\pi}{2}\sigma_x)\otimes \exp(\I \frac{\pi}{2}\sigma_x)$ and $\exp(\I \frac{\pi}{2}\sigma_y)\otimes \exp(\I \frac{\pi}{2}\sigma_y)$ respectively. These are combined to estimate the singlet state fidelity, using $F(\ket{\Psi^{-}}) = \frac{1}{4} (1 - \langle \sigma_{x} \sigma_{x}\rangle - \langle \sigma_{y}\sigma_{y}\rangle - \langle \sigma_{z} \sigma_{z}\rangle)$. 

\begin{figure}[ht]
\includegraphics{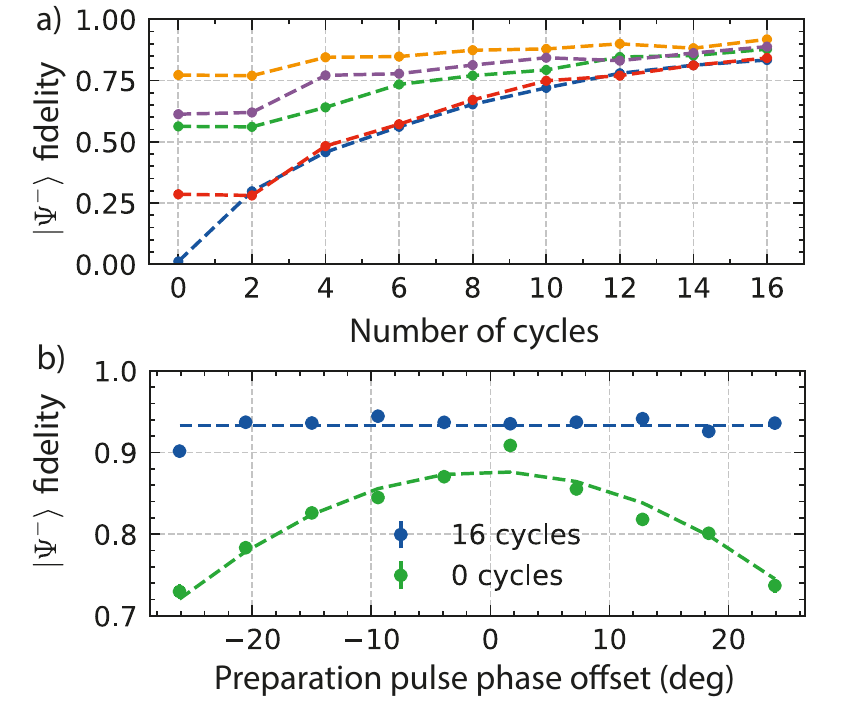}
\caption{(a) The effect of applying up to 16 cycles of the protocol to various initial states. All cases converge towards $\ket{\Psi^{-}}$. Fidelity differences between different input states  decay exponentially with the number of cycles, but remain resolvable after 16 cycles (b) Comparison of the singlet fidelity before and after the protocol is applied. Different initial states are generated by changing the phase of the $\ket{\Psi^{-}}$ preparation pulse (Supp. Mat.~\ref{sec:experiment_details}), which also simulates the effect of a laser frequency error. After 16 cycles, states with fidelities $\gtrapprox 0.75$ converge onto the same steady state. Error bars, corresponding to $\pm 1\sigma$ confidence intervals, are smaller than data points.}
\label{fig:data}
\end{figure}
Figure~\ref{fig:data} a) shows the measured fidelity as a function of the number of cycles of the protocol, applied to a range of initial states, showing the expected convergence towards the singlet. Different starting states were created by initializing the qubit to $\ket{\downarrow\downarrow}$ and mapping it to a mixture of singlet and triplet states (Supp. Mat.~\ref{sec:experiment_details}), with the motional mode used for the collective excitation cooled close to the ground state. Figure~\ref{fig:data} b) illustrates how, after 16 cycles of the protocol, states with initial fidelities $\gtrapprox 0.75$ are mapped onto output states with the same final fidelity. Averaged over all the data, we find a fidelity of $93(1)\%$ at 16 cycles. 

\begin{figure*}
\includegraphics{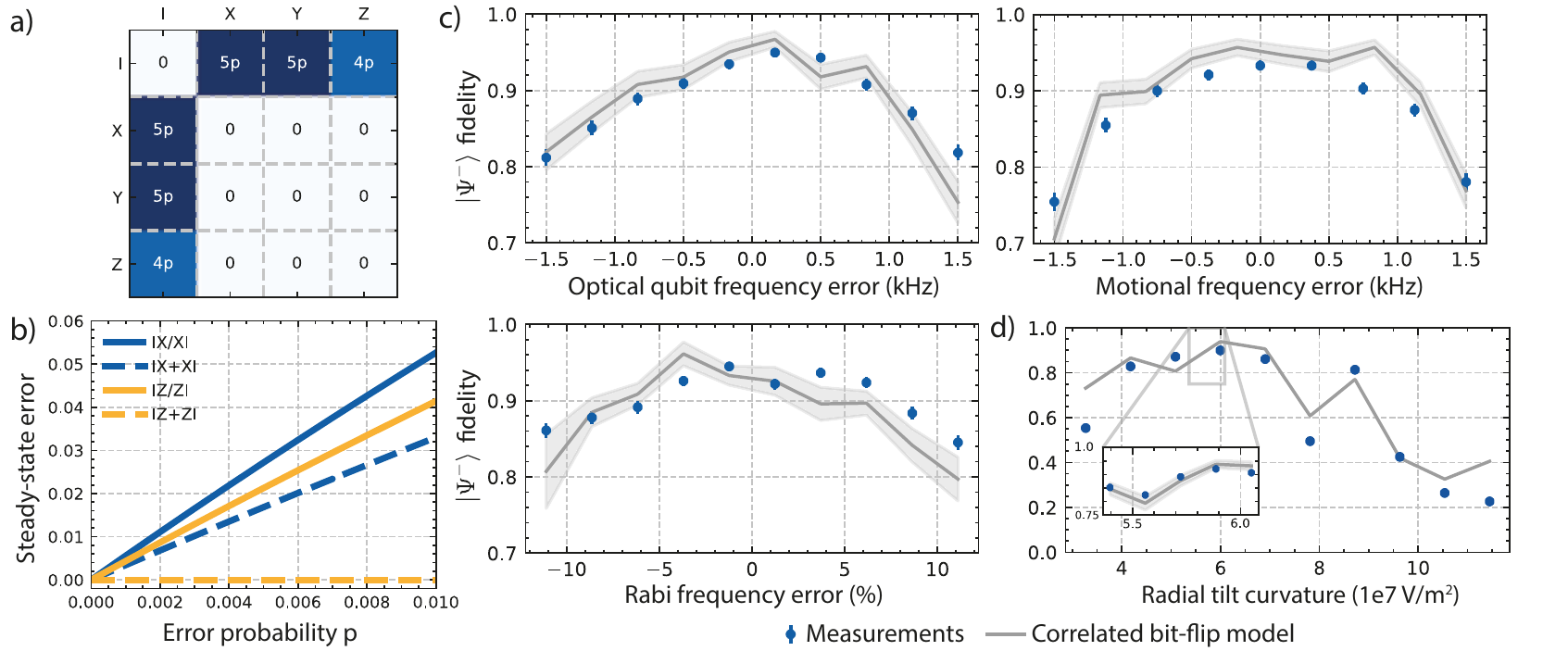}
\caption{a) The steady-state error associated with individual error channels of probability $p$. b) Comparison of steady-state errors associated with uncorrelated (solid) and correlated (dashed) errors. c,d) Experimentally measured values of $F(\ket{\Psi^{-}})$ (blue dots) compared to the prediction ($1-3.2p$) of a correlated bit-flip error model (gray lines), with $p$ obtained from independent experimental measurements. Error bars show a $\pm 1 \sigma$ confidence interval, but are smaller than most data points. In several data points in d) the measured values of $p$ are large, and we can no longer apply the linear approximation.}
\label{fig:errors}
\end{figure*}

We analyze the noise robustness of the protocol by considering the bichromatic drive (A) as the dominant source of errors. Assuming $\ket{\uparrow}$ is spectroscopically decoupled from (A), we can describe all errors through 16 elementary error channels $\{I_e,X_e,Y_e,Z_e\}^{\otimes 2}$ acting on the $\{\ket{\downarrow},\ket{e}\}^{\otimes 2}$ subspace. The effect of those errors acting with probability $p$ per cycle in a depolarising model is shown in Fig~\ref{fig:errors} a) (see Supp. Mat. \ref{sec:error_channels} for precise definitions).

We find that the final fidelity is independent of all global errors (such as $X_e X_e$ or $X_e Z_e$). On the other hand, all local errors (such as $X_e I_e$ or $I_e Z_e$) become amplified. A particularly experimentally relevant class of errors is \textit{correlated local errors}. These include correlated bit-flip errors (corresponding to an application of the operator $I_eX_e+X_eI_e$) which arise due to residual spin-motion entanglement at the end of the collective excitation step (i.e. when $\alpha(t) \neq 0$) or off-resonant excitation of spectator transitions. Magnetic-field fluctuations common to both ions would produce a correlated phase-flip error ($I_e Z_e+Z_e I_e$). We find that such correlations increase the fidelity of the collective optical pumping compared with uncorrelated errors with similar constituent operators. Results of simulations showing this are displayed in Fig~\ref{fig:errors} b). For example, a bit-flip error with probability $p$ per cycle reduces the singlet fidelity by $\approx 5.2p$ when uncorrelated and $\approx 3.2p$ when correlated (Supp.Mat~\ref{sec:error_channels}). Correlated phase-flip errors leave the fidelity unaffected since $\ket{\Psi^{-}}$ resides in a decoherence-free subspace.

These insights are matched by simulations of the dynamics of the collective optical pumping in the presence of experimentally relevant error sources. These reveal that, compared to either a single entangling gate or a two-loop phase-modulated entangling gate \cite{Milne2020} based on the Hamiltonian $H_A$, our protocol reduces the effect of qubit frequency errors and Rabi frequency errors. For motional frequency errors and fast (Markovian) optical qubit dephasing, our protocol does not provide benefits. Details and discussion of these results are presented in Supp. Mat. \ref{sec:error_sensitivity}. 



It is challenging to exactly account for the measured error from first principles due to a number of setup-specific imperfections. We experience \si{\kHz}-level drifts in motional mode frequencies due to charging of the trap surface caused by shining light through the integrated waveguides \cite{Harlander2010a}. Following the initial transients, these fluctuations alone are too slow to significantly affect the collective excitation modelled as in Eq.~\eqref{eq:MS_unitary_general}. However, the  drifts in spectator mode frequencies occasionally lead to a mode spectrum where the collective excitation drive (A) off-resonantly excites spectator optical transitions. Once large, this error can be corrected by tuning mode frequencies, but it is challenging to estimate its magnitude between calibrations. These drifts were the primary reason we worked with a fixed number of protocol cycles ($N=16$, corresponding to $\approx \SI{3}{\milli\second}$ of $\SI{729}{\nm}$ light per shot). High heating rates mean that each cycle of the protocol starts with a higher occupancy of the spectator modes ($\approx 0.5$ quanta per cycle on a \SI{1}{\MHz} center-of-mass mode), leading to an increase in a correlated bit-flip error during the drive (A) as the protocol progresses. We measure that a bit-flip error probability of $p \approx 0.01$ per cycle increases to $p \approx 0.02$ after 16 cycles. The measured 16-cycle error of $0.07(1)$ is consistent with the value of $3.2p$ predicted by the correlated bit-flip error model. The obtained fidelity is significantly reduced compared to the unitary gate based on the same Hamiltonian $H_A$, which produces $\left(\ket{\downarrow \downarrow} - i \ket{ee}\right)/\sqrt{2}$ with fidelity of $\gtrapprox 99\%$ \cite{Mehta2020}, though it does improve our ability to prepare $\ket{\Psi^{-}}$, which is currently limited by errors in single-ion addressing in our coherent implementation.

Instead of comparing the measured error with \emph{a priori} simulations, we measure the 16-cycle fidelity, as well as the bit-flip probability in the collective excitation step, for a range of experimental miscalibrations. We mitigate motional mode drifts by keeping the average applied power of the \SI{729}{\nm} laser constant during the data acquisition period. For each parameter, we experimentally approximate the steady-state fidelity $F(\ket{\Psi^{-}})$ by first preparing $\ket{\Psi^{-}}$ with fidelity around $0.75$ using unitary methods, and then applying 16 cycles of the protocol. The bit-flip probability $p$ for the last cycle is independently estimated by applying 16 cycles of the protocol, followed by optical pumping and a single round of drive (A) (Supp. Mat. \ref{sec:experiment_details}). The comparison between $F(\ket{\Psi^{-}})$ and the bit-flip model prediction is shown in Figure~\ref{fig:errors} c). We find qualitative agreement, suggesting that the bit-flip model accurately captures the errors of the protocol. Figure~\ref{fig:errors} d) illustrates the challenge associated with spectral crowding. We modify the spectator mode spectrum by adding an additional quadrupole potential with eigenaxes at $\pm \SI{45}{\deg}$ to the trap surface in the radial plane. This changes the radial spectator mode orientations, frequencies and temperatures, while keeping the (axial) gate mode frequency approximately constant. We find that the spectator spectrum is clear only for a narrow range of curvatures (here between $\SI{5.7e7}{\volt\per\metre^2}$ and $\SI{6.1e7}{\volt\per\metre^2}$), which needs re-calibrating every few hours. Further qualitative discrepancies between the results in Fig~\ref{fig:errors} c) and the simulations in Supp. Mat. \ref{sec:error_sensitivity} can be accounted for by the $\approx 2 \pi \times \SI{17}{\kHz}$ dipole AC Stark shift of the drive (A), combined with a gap time between the two segments of the drive (A) (Supp. Mat. \ref{sec:experiment_details}). 

All of the limitations listed above are setup specific and do not pose a fundamental limitation to the protocol. Coupling to spectator transitions could be suppressed by increasing the magnetic field or improving the spectrum of the laser. The heating rate observed in this trap is particularly high, exceeding levels observed in cryogenic traps with comparable ion-electrode distances by a factor of $\approx 100$ \cite{Brownnutt2015, Sedlacek2018}. Reducing it to more typical levels, combined with better shielding of nearby dielectrics \cite{Hong2018, Ragg2019}, would suppress drifts within each collective pumping sequence and hence allow the protocol to reach its true steady state. Alternatively, motional mode temperature could be stabilized throughout the protocol by sympathetic cooling \cite{Home2008}. Finally, by compensating the dipole AC Stark shift with an additional tone of the \SI{729}{\nm} laser \cite{Haffner2003}, one could reduce the sensitivity to Rabi frequency and optical qubit frequency errors.

We have presented and implemented a novel protocol for collective optical pumping into a maximally entangled two-qubit state. We measure a singlet fidelity of $93(1)\%$ after 16 cycles of the protocol (at which point a quasi steady-state has been achieved), to our knowledge higher than any dissipative method to date. The observed infidelity is consistent with measured bit-flip errors of the effective $S_x^2$ drive which for our implementation is mediated by a motional mode. The protocol can be practically beneficial in experiments limited by global errors, especially as a method of purifying lower-fidelity Bell states. Dissipative generation of high-fidelity entangled states could find application in a variety of quantum information processing tasks, such as dissipative encoding \cite{Baggio2021}, error-corrected quantum sensing \cite{Rojkov2021}, and as a supply of entangled resource states for quantum gate teleportation \cite{Gottesman1999}. 

While the analysis in this paper focused on a specific implementation in ${}^{40}\text{Ca}^+$ ions, the protocol is general, and we anticipate it might be applied in a wide range of platforms where suitable couplings are available. In nitrogen-vacancy centers, collective excitation may be achieved by direct spin-spin interactions \cite{Maze2011}. In neutral atom platforms, Rydberg dressing provides a mechanism suitable for implementing the collective excitation \cite{Mitra2020}. Furthermore, parametrically-driven superconductors allow for the realization of two-qubit interactions for dissipation engineering \cite{Doucet2020}.


\begin{acknowledgments}
We acknowledge funding from the Swiss National Science Foundation (grant no. 200020$\_$165555), the National Centre of Competence in Research for Quantum Science and Technology (QSIT), ETH Z\"urich, and the Intelligence Advanced Research Projects Activity (IARPA) via the US Army Research Office grant W911NF-16-1-0070. FR and IR acknowledge financial support from the Swiss National Science Foundation (Ambizione grant no. PZ00P2$\_$186040).

\textbf{Author Contributions:} JPH devised the scheme, which was then simulated and analyzed by MM and VN. MM, CZ and KKM carried out the experiments presented here in an apparatus with significant contributions from MM, CZ, KKM, TLN, and MS. MM analyzed the data and constructed the error model. IR and FR developed an analytic approach to describe the protocol. VN performed an initial experimental investigation. JPH, KKM and DK supervised the work. The manuscript was written by MM, FR, and JPH with inputs from all authors.

\end{acknowledgments}


\bibliography{main}

\iftoggle{arXiv}{\iftoggle{arXiv}{
	\clearpage 
	\begin{center}
	{\LARGE\bfseries Supplementary Material}    
	\end{center}
}{}
\section{Convergence rates}
\label{sec:convergence_rate}
For the calculations and convergence rate simulations, we write a single cycle of the protocol as a superoperator
\begin{align}
    S(\Phi,\gamma,\theta) = S_C(\theta) S_B(\gamma) S_A(\Phi),
\end{align}
where $S_A(\Phi),S_B(\gamma)$ and $S_C(\theta)$ are the superoperators for drives (A), (B), and (C) respectively. $S_A$ and $S_C$ are obtained by converting unitary operators $U_A$ and $U_C$ in the main text to superoperators. Drive (B) is a composition of two independent decay channels. We describe the single-ion decay with a set of three Kraus operators $E_0 = \ket{\downarrow}\bra{\downarrow} + \ket{\uparrow}\bra{\uparrow}$, $E_1 = \sqrt{p_{e\rightarrow \downarrow}}\ket{\downarrow}\bra{e}$ and $E_2 = \sqrt{p_{e\rightarrow \uparrow}}\ket{\uparrow}\bra{e}$. The total decay channel is then modelled as nine two-ion maps $E_i \otimes E_j$, with $i,j\in\{1,2,3\}$.

The state of the system after $N$ cycles of the protocol is then given by 
\begin{align}
    \vec{\rho}_N = \Big[ S(\Phi,\gamma,\theta) \Big]^N\,\vec{\rho}_0 \, ,
\end{align}
where $\vec{\rho}_N$ and $\vec{\rho}_0$ are the vector representation of the final and initial density matrices, respectively. Similarly to the spectral decomposition of Liouvillians~\cite{Minganti2018}, we perform an eigenanalysis of $S(\Phi,\gamma,\theta)$. The steady state of the system is the $+1$ eigenvector of the superoperator, which corresponds to $\rho_{ss}=\ket{\Psi^-}\bra{\Psi^-}$. In order to assess the convergence rate of the fidelity, we must determine the second largest eigenvalue of the the superoperator $\lambda_{\max}$. The fidelity will then take the following form
\begin{align} \label{eq:fidelity_discrete}
    F(N) \approx 1 - C_0 \, \lambda_{\max}^{N} = 1 - C_0 \, e^{-N/N_0} \, ,
\end{align}
where $C_0$ is the initial population out of singlet state and $1/N_0$ the convergence rate in unit of inverse cycles.

To simplify the analysis, we make the approximation of a full transfer during the step (A) (i.e. $\Phi = \pi/4$) and a complete repump during (B) (i.e. $S_B(\gamma)\ket{ee}$ lays in the ground "spin" state manifold). This allows us to combine drives (A) and (B) into one superoperator $S_{BA}(\gamma)$, which acts on states within the $\{ \ket{\downarrow\downarrow},\,\ket{\uparrow \uparrow},\,\ket{\Psi^+},\,\ket{\Psi^-}\}$ manifold only.

The eigenanalysis of the remaining two-parameter superoperator shows only six non-zero eigenvalues, among which $\lambda_{ss}=+1$ and $\lambda_{1,2,3,4,5}$. The latter five values are solutions of a polynomial equation of degree five which has, for most values of $\theta$ and $\gamma$, two complex, one negative and two positive roots. The desired $\lambda_{\max}$ is the maximum of the latter two. We find that $\lambda_{\max}$ strictly dominates all the other eigenvalues over the entire parameter space. Fig.~\ref{fig:discrete_eigenvalue} shows its value as a function of $(\theta,\gamma)$. The optimal convergence is attained when $\lambda_{\max}$ is minimal. Numerically, we determine that the optimal parameters are then $\theta^\mathrm{opt}\approx0.72\pi$ and $\gamma^\mathrm{opt}\approx0.22\pi$. In this configuration, the protocol converges at a rate of $1/N_0^{\mathrm{opt}}=-\log(\lambda_{\max}^\mathrm{opt})\approx 1/7.62$ per cycle.

\begin{figure}[t]
\centering
\includegraphics[trim=5pt 5pt 5pt 5pt, clip]{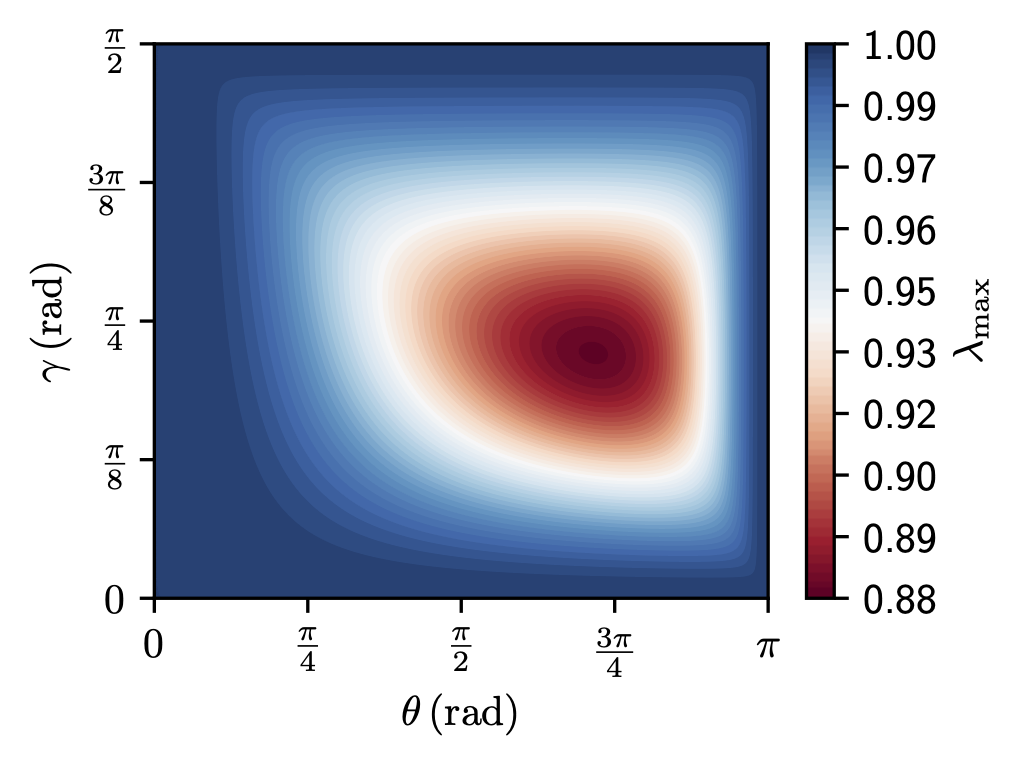}
\caption{Second dominant eigenvalue $\lambda_{\max}$ after $\lambda_{ss}=+1$ of the superoperator $S(\pi/4,\gamma,\theta)$ as a function of protocol's parameters $\gamma$ and $\theta$. This value sets the convergence rate of the fidelity as described in Eq.~\eqref{eq:fidelity_discrete} and is optimal for ${\theta^\mathrm{opt}\approx0.72\pi}$ and $\gamma^\mathrm{opt}\approx0.22\pi$}
\label{fig:discrete_eigenvalue}
\end{figure}

To appreciate the general convergence behavior of our method, we simulate an ideal process for an initial state $\ket{\downarrow\downarrow}$, and calculate the final overlap with $\ket{\Psi^-}$. Fig.~\ref{fig:convergence} a) shows how the singlet error after 50 cycles depends on the protocol parameters. The convergence slow-down around $\Phi=m \pi/2$ corresponds to a very low $\ket{\downarrow\downarrow}\rightarrow\ket{ee}$ state transfer probability per cycle. The slow convergence around $\gamma=m \pi/2$ corresponds to very uneven branching ratios, resulting in $\ket{ee}$ predominantly decaying to either $\ket{\downarrow\downarrow}$ or $\ket{\uparrow\uparrow}$. Finally, $\theta = m \pi$ results in population trapping in $\ket{\Psi^{+}}$ for odd $m$, while for even $m$, the population is trapped in $\ket{\uparrow \uparrow}$ as well. Fig.~\ref{fig:convergence} b) illustrates how much a large error in $\theta$ slows down the convergence.
\begin{figure*}
\includegraphics{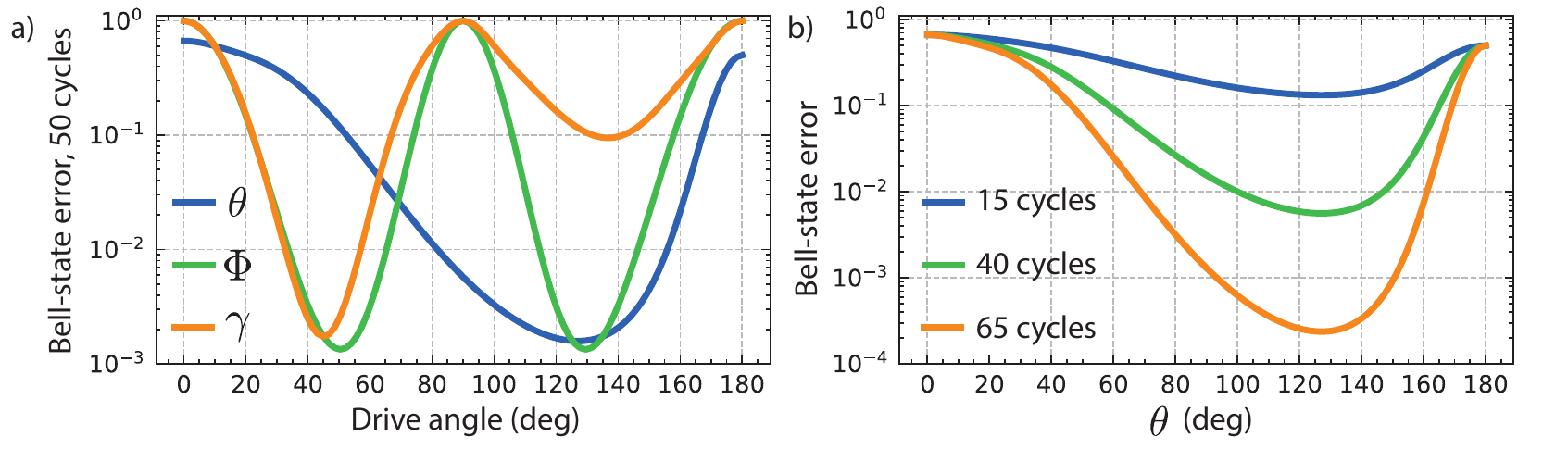}
\caption{a) Error after 50 cycles of the protocol for different values MS angle $\Psi$, branching angle $\gamma$ and the ground-state drive angle $\theta$. In each line, one parameter is varied, and the others are kept at $(\Psi,\gamma,\theta) = (\frac{\pi}{4},\frac{\pi}{4},\frac{3\pi}{4})$. b) Error variation vs $\theta$ for increasing number of cycles.}
\label{fig:convergence}
\end{figure*}

As mentioned in the main text, due to the difference in the differential AC Stark shift we must switch to sequential $\theta_1=\pi$ and $\theta_2=\frac{\pi}{2}$ drives (C) such that the superoperator that we must spectrally decompose is now $S(\pi/4,\gamma,\theta_2) S(\pi/4,\gamma,\theta_1)$. This particular choice of even and odd drives drastically simplifies the system, reducing the number of non-zero eigenvalues to three. These are $\lambda_{ss}=+1$ and
\begin{align} \label{eq:eigenval_alternate_cycles}
    \lambda_\pm = \frac{1}{4} \left( 1 \pm 3 \sqrt{ 1 - \frac{2}{9} \,\Big[2 + \cos^4(\gamma)\Big]\,\sin^2(2\gamma)} \right) \, .
\end{align}
The converence rate is then set by $\lambda_{+}$. However, to compare it to $\lambda_{\max}$ calculated for a sequence of constant drives (C), we must recall that $\lambda_{+}$ is now the eigenvalue of a composition of two cycles, an odd and an even one. Since the slowest process in the protocol is the drive (A), we can assume that $S(\pi/4,\gamma,\theta_1)$ and $S(\pi/4,\gamma,\theta_2)$ have approximately the same duration and thus $\lambda_{\max}$ is fairly comparable to $\lambda_{+}^{1/2}$. Fig.~\ref{fig:lambda_const_vs_altern} depicts this comparison as a function of the branching ratio angle $\gamma$.

\begin{figure}[b]
\vspace{-10pt}
\includegraphics[trim=5pt 7pt 5pt 10pt, clip]{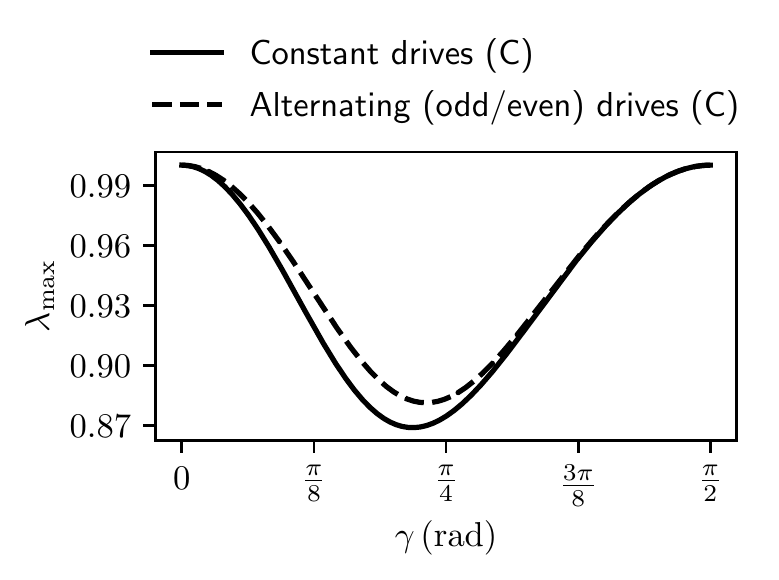}
\caption{Comparison of the second dominant eigenvalue $\lambda_{\max}$ which determines the converge of the singlet state fidelity (cf. Eq.~\eqref{eq:fidelity_discrete}) for scenarios with constant drives (C) for all cycles and with alternating drives for odd and even cycles. In the latter case, we have chosen $\lambda_{\max} = \lambda_{+}^{1/2}$ where $\lambda_{+}$ defined in Eq.~\eqref{eq:eigenval_alternate_cycles} is the second dominant eigenvalue of $S(\pi/4,\gamma,\theta_2) S(\pi/4,\gamma,\theta_1)$ which is approximately twice as long as one simple cycle.}
\label{fig:lambda_const_vs_altern}
\end{figure}

In the scheme with alternating drives (C), the protocol converges optimally when $\gamma^\mathrm{opt}\approx0.23\pi$ and it does it with a rate of $1/N_0^{\mathrm{opt}}=-\frac{1}{2}\log(\lambda_{+}^\mathrm{opt})\approx 1/7.92$ per cycle. For the experimentally measured $\gamma\approx0.31\pi$, the convergence rate equals $1/10.35$ per cycle.

\section{Discrete error channels}
\label{sec:error_channels}
The errors which act on the levels $\ket{\downarrow}$ and $\ket{e}$ (involved in the collective excitation), and not on $\ket{\uparrow}$, correspond to evolution under a tensor product of two Pauli operators acting on the $\{\ket{\downarrow},\ket{e}\}$ subspace. The resulting 16 error operators can be defined as:
\begin{align}
    M_{i,j} = \Exp{-\I (\pi/2) \sigma_{i,\downarrow e} \otimes \sigma_{j,\downarrow e}}
\end{align}
where $\sigma_{i,\downarrow e}$ is a Pauli matrix on the $\{\ket{\downarrow},\ket{e}\}$ subspace, $i,j \in \{0,x,y,z\}$, using $\sigma_0=I$. This operator definition extends the usual notion of single-qubit errors onto a three-level system. We define an error operator $M_{i,j}$ as local when $i=0$ or $j=0$, and as global otherwise. To study the influence of each of these operators individually, we simulate the effect of introducing a map
\begin{align}
    \rho \rightarrow (1-p) \rho + p M_{i,j} \rho M_{i,j}^\dagger
\end{align}
acting once per cycle after drive (A). The resulting error probabilities in Fig.~\ref{fig:errors} a) are obtained by fitting the numerical results for $p<0.01$, and are rounded to the nearest integer. For larger error probabilities, the error is nonlinear is $p$.

We address correlated errors with two methods. Correlated errors can arise due to undesired coherent processes, such as coupling to spectator modes. We can model such correlated bit-flips and phase-flips as undesired x/z rotations of the optical qubit. To that end, we introduce in the simulation two-qubit rotation operators
\begin{align}
    U_i(\epsilon) = \Exp{\I (\epsilon/2)\sigma_{i, \downarrow e}}\otimes \Exp{\I (\epsilon/2)\sigma_{i, \downarrow e}}
\end{align}
where $i=x$ corresponds to a correlated bit-flip and $i=z$ corresponds to a correlated phase-flip.
This operator is applied once per cycle, after the drive (A). If both ions are in the $\{\ket{\downarrow},\ket{e}\}$ subspace, this creates an error with probability $p \approx 2 (\frac{\epsilon}{2})^2$ for $p \ll 1$. This allows us to parametrize this process with $\epsilon = \sqrt{2 p}$. In the main text, the correlated bit-flip error ($I_eX_e+X_eI_e$) is modelled by $U_{x}(\sqrt{2 p})$ and the correlated phase-flip error ($I_eZ_e+Z_eI_e$) is modelled by $U_{z}(\sqrt{2 p})$.

Residual spin-motion entanglement likewise results in correlated bit-flip errors, although they are not represented by unitary operations. We can model such errors as a set of Kraus operators corresponding to different final occupations of the oscillator. In the limit of $|\alpha(t)| \ll 1$, any residual displacement $\alpha(t)$ creates a correlated bit-flip with probability $p=|\alpha(t)|^2$, corresponding to an undesired change of the motional state by 1 quantum. The Kraus operator corresponding to a correlated bit-flip is then $I_eX_e+X_eI_e$. In simulations, we find that the steady-state error caused by this residual spin-motion entanglement is almost identical to the steady-state error introduced by $U_{x}(\sqrt{2 p})$. Thus, we simulations presented in the main text apply identically to both of these errors sources.

Thanks to these definitions, the value of $p$ can be estimated for both the correlated and uncorrelated bit-flip error by preparing the system in $\ket{\downarrow\downarrow}$ and applying a single cycle of the drive $A$. Then, $p$ is equal to the probability to find exactly one ion in $\ket{\downarrow}$. With this parametrization, the correlated bit-flip error produces a lower steady-state singlet error than an uncorrelated bit-flip error. However, it does not mean that the same conclusion holds in the Markovian limit for correlated and uncorrelated bit-flip occurring at the same rate (Supp. Mat.~\ref{sec:continuous_scenario}), as the difference may be an artefact of the parametrisation.

\section{Simulations of error sensitivity}
\label{sec:error_sensitivity}
We estimate the error sensitivity of the protocol in the trapped-ion implementation using a simulation that includes the full Hamiltonian for the drive (A). When multiple pulses are applied in a sequence, it is essential to control the phases of the sideband drives. Consider drive (A) with blue and red sidebands phases $\phi_b = (\phi_s + \phi_m)/2$ and $\phi_r = (\phi_s - \phi_m)/2$ respectively. In a presence of a qubit frequency error $\epsilon_q$ and a motional frequency error $\epsilon_m$, the interaction Hamiltonian for drive (A) becomes:
\begin{equation}
    \label{eq:ms_hamiltonian}
    H_A = \frac{\eta \hbar \Omega}{2} S_{e,\phi} (\hat{a} \Exp{\I (\delta+\epsilon_m+\epsilon_q) t}\Exp{\I \phi_m} + \hat{a}^\dagger  \Exp{-\I (\delta+\epsilon_m-\epsilon_q) t}\Exp{-\I \phi_m}),
\end{equation}
where $\eta$ is the Lamb-Dicke parameter, $\Omega$ is the carrier Rabi frequency, $\hat{a}$ is the oscillator annihilation operator, and the total spin operator is a sum of two single-ion operators $S_{e,\phi} = S_{e,\phi}^{(1)} + S_{\phi}^{(2)}$, where
\begin{align}
    S_{e,\phi} = \ket{e}\bra{\downarrow} \Exp{\I \phi_s} +  \ket{\downarrow}\bra{e} \Exp{-\I \phi_s}.
\end{align}
In writing down Eq.~\eqref{eq:ms_hamiltonian}, we took a rotating-wave approximation to remove components oscillating faster than $\delta$, and thus implicitly assumed that the $\ket{\downarrow}\rightarrow\ket{e}$ transition and its sidebands are spectroscopically decoupled from the MS interaction. Eq.~\eqref{eq:ms_hamiltonian} reduces to $H_A$ in the main text for $\phi_s=\phi_m=0=$ and $\epsilon_m=\epsilon_q=0$.

We simulate each cycle as a sequence of two pulses of length $t$. We set $\phi_{s1}=\phi_{m1}=0$ for the first pulse while the second has $\phi_{s2}=0$, $\phi_{m2}=\pi - \delta t$. The simulation parameters are: $\delta = 2\pi\times\SI{15}{kHz}$, $t=2 \pi\delta = \SI{66.6}{\micro\second}$, $\eta = 0.028$, $\Omega = \delta/(2 \eta) = 2 \pi \times \SI{265}{\kHz}$. For comparison, we also simulate a single-loop MS gate (total length $t$) and a two-loop MS gate (total length $t\times\sqrt{2}$) which are based on the same Hamiltonian $H_A$. For all the simulations, the state is initialised in $\ket{\downarrow\downarrow}$ and in the motional ground state. For our protocol, we extract the fidelity with $\ket{\Psi^-}$ after 80 cycles, while for the MS gates, we calculate the fidelity between the final state and $(\ket{\downarrow\downarrow}-\I\ket{ee})/\sqrt{2}$. 

\begin{figure*}
\includegraphics[width=1\linewidth]{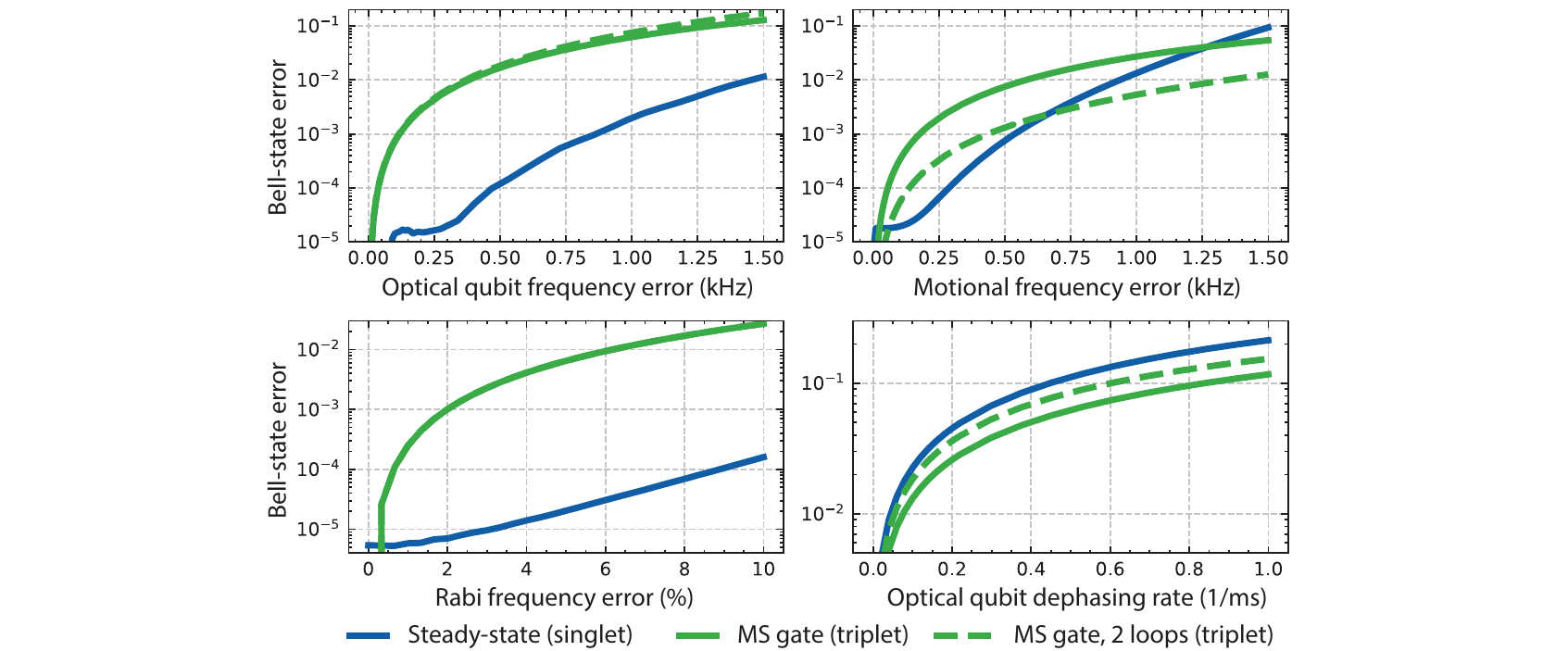}
\caption{Comparison of the Bell-state error of an MS gate (green, solid), phase-modulated two-loop MS gate (green, dashed) and 80 cycles of our protocol (blue) applied to an initial state $\ket{\downarrow\downarrow}$. Note that the ideal Bell state is a spin triplet for the MS gates, and a spin singlet for our protocol. Optical qubit frequency error (top left) is simulated by setting $\epsilon_q > 0$ in Eq.~\eqref{eq:ms_hamiltonian}. Motional frequency error (top right) is simulated by setting $\epsilon_m>0$ in Eq.~\eqref{eq:ms_hamiltonian}. Fractional Rabi frequency error $\epsilon_\Omega$ (bottom left) is simulated by setting $\Omega=(1+\epsilon_\Omega)\delta/(2 \eta)$ in Eq.~\eqref{eq:ms_hamiltonian}. Error caused by optical qubit dephasing at rate $\Gamma$ (bottom right) is simulated by solving the Lindblad master equation with Hamiltonian in Eq.~\eqref{eq:ms_hamiltonian} and a jump operator $L=\sqrt{\Gamma}(\sigma_{z,\downarrow e}^{(1)} + \sigma_{z,\downarrow e}^{(2)})$ appropriate for correlated phase noise.
}
\label{fig:error_simulations}
\end{figure*}
Fig.~\ref{fig:error_simulations} shows the simulations of the sensitivity to experimental errors. The reduced sensitivity to optical qubit frequency offsets can be understood as follows. The first-order effect of a qubit frequency error on an MS gate is that it affects the phase of the final entangled state (i.e. the final state is close to $(\ket{\downarrow\downarrow}+\Exp{\I \phi}\ket{ee})/\sqrt{2}$, but $\phi$ varies). Our protocol is insensitive to correlated phase errors and therefore achieves error suppression. Experimentally, the optical qubit frequency errors can be caused by drifts in magnetic fields or the qubit laser frequency (since the optical qubit frequency is specified in the frame of the laser). 

The protocol is also beneficial for Rabi frequency errors which correspond to unwanted $X_e X_e$ processes. The simulated error increase for 80 cycles is solely a result of a slowdown in convergence -- the steady-state error is completely unaffected.

Our method is not particularly beneficial in the presence of gate mode frequency errors, which can be caused by drifting trap potentials. Such an error causes two effects of comparable magnitude on a single MS gate. First, the gate area is modified, corresponding to an undesired $X_e X_e$ process. Second, the loop does not close, resulting in residual spin-motion entanglement and hence a correlated bit-flip. Our protocol removes the first error but amplifies the second, and thus the overall error stays comparable. While the simulations presented in Fig.~\ref{fig:error_simulations} indicate an error suppression for small frequency errors, we note that this could be achieved equally well (but in a much shorter time) by using higher-order phase modulation.

Likewise, we find no benefit of using our protocol if an experiment is limited by fast (Markovian) optical qubit dephasing, which could be caused by fast laser or magnetic field noise. This is because dephasing during the MS gate causes a bit-flip error with probability $p$ proportional to the gate time $t$. 

An important effect not included in the simulation is the spontaneous emission from state $\ket{e}$. The optical qubit lifetime $T_1$ limits the Bell state error of an MS gate of length $t$ to $\approx 0.5 t/T_1$. The resulting bit-flip error becomes amplified by the protocol, resulting in a lower-bound on the steady-state infidelity of $\approx 3t/T_1$.

Finally, the entangling drive (A) may not be the only source of errors in the protocol. The dissipative step (B) needs to be sufficiently spectroscopically decoupled to avoid exciting the ground state qubit. Besides this, it is only necessary to ensure the population does not get trapped outside of the protocol subspace. Ultimately, every dissipation event can cause heating due to photon recoil, but the average number of photon recoil events will be significantly lower than the number of cycles. For a branching ratio parametrized by $\gamma$, the average number of photons scattered during the full protocol $n_{\gamma}=4\csc(2\gamma)^2$. For $\gamma=0.31\pi$ this gives $n_{\gamma}\approx 4.5$ photons.  Thus, photon scattering during the protocol produces negligible heating, comparable to this caused by optical pumping.

The most important assumption about the drive (C) is that it drives both ground-state qubits with equal Rabi frequencies. For a small imbalance of $\Omega_C^{(2)} = \Omega_C^{(1)}(1+\epsilon), \epsilon \ll 1$ and $\theta = 3 \pi/4$, the singlet state gets repumped with probability $p \approx 1.4 \epsilon^2$ per cycle. This increases the steady-state error similarly to a correlated bit-flip during the drive (A). Obtaining $p \ll 10^{-4}$ requires $\epsilon \ll 10^{-2}$, which can be easily the case for long-wavelength radiation (although we achieve $\epsilon < 10^{-2}$ for the optical qubit as well). Magnetic field noise does not impact the steady-state fidelity, since $\ket{\Psi^{-}}$ lives in a decoherence-free subspace, but noise in magnetic field gradient does cause coupling between $\ket{\Psi^{-}}$ and $\ket{\Psi^{+}}$. Off-resonant couplings of drives (A)-(C) to spectator modes (AC stark shifts) generally do not cause an error, since they affect $\ket{\downarrow}$ and $\ket{\uparrow}$ equally. Furthermore, any drives that affect the Zeeman qubit states differentially (from the \SI{729}{\nm} laser) are still expected to couple to both ions equally, and thus leave $\ket{\Psi^{-}}$ unaffected. The exception is presented in the next section.

\section{Differential phase shift cancellation}
\label{sec:differential_phase_shift}
In the process of performing the experiments, we discovered that drive (A) causes a coupling between $\ket{\Psi^{-}}$ and $\ket{\Psi^{+}}$. Further investigation revealed the following explanation. For the protocol, we position the ions such that their carrier Rabi frequencies on the $\ket{\downarrow} \rightarrow \ket{e}$ transition are equal ($<1\%$ difference). However, to our surprise, when we drive different carrier transition between $S_{1/2}$ and $D_{5/2}$ states at the same position, we observe significant Rabi frequency gradients. Most importantly, we find a Rabi frequency difference of $\epsilon \approx 5\%$ on a carrier transition $\ket{\uparrow}\rightarrow\ket{D_{5/2},m_J = +3/2}$ which is detuned by $ \approx 2 \pi \times \SI{0.9}{\MHz}$ from the blue sideband of the axial stretch mode. When drive (A) is applied, the blue sideband tone causes an AC Stark shift of $\Delta \approx 2 \pi \times \SI{25}{\kHz}$ to be applied  on $\ket{\downarrow}$. Because of the Rabi frequency difference, the states $\ket{\downarrow \uparrow}$ and $\ket{\uparrow\downarrow}$ acquire a relative phase of $\phi = 2 \epsilon \Delta t$ after time $t$. For a system initially in $\ket{\Psi^{-}}$, the probability to find it in $\ket{\Psi^{+}}$ after time $t=2\times \frac{2\pi}{\delta}$ (a single cycle of drive (A)) is given by $p=\sin^2(\phi/2) \approx 0.5$. Such a large error probability per cycle completely destroys the protocol.

Fortunately, this coupling is coherent and can be fully reversed by flipping the ground state spins mid-way through the evolution. In order not to affect the performance of the drive (A), we do not insert the spin-echo pulse in-between its two segments. Instead, we turn every odd Zeeman drive (C) into a $\pi$-pulse. This way, if the population is in $\ket{\uparrow\uparrow}$ it gets transferred to $\ket{\downarrow\downarrow}$, while if it is in the $\ket{\Psi^{\pm}}$ manifold, the subsequent application of the drive (A) reverses the previously acquired phase error $\phi$. This way, at the end of each even cycle, the population returns to $\ket{\Psi^{-}}$. In the even cycles, we apply a Zeeman $(\pi/2)$-pulse to move the population from $\ket{\Psi^{+}}$ to $\ket{\downarrow\downarrow}$. 

The origin of $\epsilon > 0$ is still being investigated. A portion of the effect could be accounted for by the non-Gaussian components in the laser mode, but we have not yet managed to construct a realistic model that fully explains the Rabi frequency differences between different carrier transitions.

\section{Experiment details}
\label{sec:experiment_details}
The experiment begins by calibrating a single MS gate with detuning $\delta = 2 \pi \times \SI{14.7}{\kHz}$. In order to avoid undesired off-resonant excitations, we slowly ramp the laser pulse on- and off, resulting in a total gate time of $t = \SI{75}{\micro\second}$, which is longer than $2 \pi/\delta = \SI{68}{\micro\second}$. Drive (A) is implemented by repeating the MS pulse twice with a motional phase shift of $\pi$ in between. Initially, we aimed to instantaneously update the phase difference between the blue and red-sideband tones in the middle of the pulse, but this produced strong off-resonant excitations, which were also confirmed in the simulation. We therefore decided instead to use two separate slowly shaped pulses, with an additional gap time of \SI{3.2}{\micro\second}. We calibrate the phases of the second MS drive to compensate for the phase offsets accumulated during the ramp- and gap time. 

The drive (C) is implemented by putting approximately \SI{18}{\milli\ampere} of current oscillating at the frequency of $2\pi\times\SI{16.5}{\MHz}$ through a loop embedded in the trap holder. The signal is generated by mixing, filtering, and amplifying two DDS sources. The signal is then sent to the cryogenic chamber, where it is band-pass filtered before entering the loop. On the trap holder, two symmetrically positioned tracks at a distance of \SI{1.1}{\mm} from the ion create an oscillating B-field of \SI{5.4}{\micro\tesla} perpendicular to the trap surface and the quantization axis. This drives a magnetic dipole transition between $\ket{\downarrow}$ and $\ket{\uparrow}$ with a Rabi frequency $\Omega_C = 2 \pi \times \SI{78}{\kHz}$ ($\pi$-time of \SI{6.4}{\micro\second}). A ground plane embedded in the trap chip protects the ion from any electric field noise from the track.

To measure the ground-state populations, we shelve $\ket{\downarrow}$ into ancillary states $\ket{D_{5/2},m_J=+3/2}$ and $\ket{D_{5/2},m_J=+1/2}$ with two \SI{729}{\nm} carrier pulses. During some data acquisition periods, we found that the latter pulse can off-resonantly drive a spectator sideband, and thus excite $\ket{\uparrow}$ (this effect varies from day to day due to changes in trap potentials). In those cases, we disabled the second shelving pulse, and instead kept track of the magnitude of shelving errors to correct them in post-processing. The measured steady-state fidelity of $93(1)\%$ is consistent between the datasets with and without the shelving error correction. After shelving, the ion state is measured by state-dependent fluorescence. The emitted photons are collected on a photo-multiplier tube (PMT), without the ability to discriminate the ions. We typically record $62$ photons per bright ion in a \SI{250}{\micro\second}. The error in distinguishing 0 bright ions from 1 bright ion is negligible, while the overlap of the 1- and 2-bright ion results in $\approx 10^{-3}$ probability of $P(\uparrow\uparrow)$ to be mislabeled as $P(\uparrow\downarrow)+P(\downarrow\uparrow)$ upon thresholding. We correct for this error in real-time by periodically measuring the histogram overlap and calculating the results as
\begin{align}
    P(\uparrow\uparrow) &= \frac{(1-p) P(2) - p P(1)}{1-p-q}\\
    P(\uparrow\downarrow) + P(\downarrow\uparrow) &= \frac{(1-q) P(2) - q P(1)}{1-p-q}
\end{align}
where $P(1)$ and $P(2)$ correspond to 1 and 2 ions recorded as bright respectively, $q$ is the probability that 2 bright ions are recorded as 1 bright, and $p$ is the probability that 1 bright ion is recorded as 2 bright.

The ground-state fidelity measurement is performed by repeating the experiment in three different settings. To extract the parity operators $\langle \sigma_{x} \sigma_{x}\rangle$ and $\langle \sigma_{y} \sigma_{y}\rangle$, we prepend the parity measurement by a Zeeman $(\pi/2)$-pulse. Since the states $\ket{\Psi^{\pm}}$ are degenerate, the $(\pi/2)$-pulse need not have an exactly specified phase. We use the fact that as long as the two $(\pi/2)$-pulses have their phases offset by $\pi/2$, the sum of the two parity measurements add up to the same value $\langle \sigma_{x} \sigma_{x}\rangle + \langle \sigma_{y} \sigma_{y}\rangle$.

To verify that the protocol indeed purifies singlet preparation errors (Fig.~\ref{fig:data}), we prepare a range of two-ion input states as follows. After ground-state-cooling the gate mode, we optically pump the system to $\ket{\downarrow\downarrow}$. We then run an MS gate followed by a of ground state mapping pulse with phase $\phi_{\text{prep}}$, which ideally prepares $\ket{\Psi^{+}}$ when $\phi_{\text{prep}}=\pi/4$. We then apply a single-ion addressing pulse of length $t_{\text{prep}}$ based on the differential AC Stark shift (Supp. Mat.~\ref{sec:differential_phase_shift}) which couples $\ket{\Psi^{+}}$ to $\ket{\Psi^{-}}$, and then run our pumping protocol. The state preparation error in is introduced either as a an offset in $t_{\text{prep}}$ (Fig.~\ref{fig:data} a) or an an offset in $\phi_{\text{prep}}$ (Fig.~\ref{fig:data} b)

\begin{figure*}[t]
    \centering
    \includegraphics{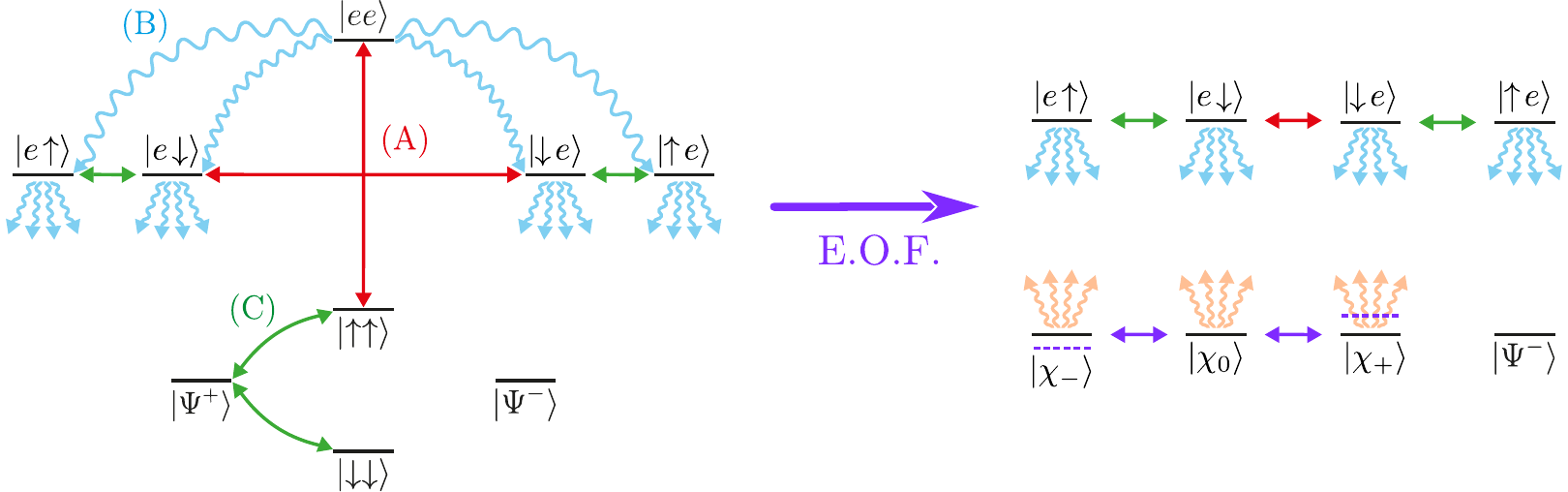}
    \caption{Energy levels and couplings of the full and effective continuous scenario. (Left) The diagram represents the full system described by the Hamiltonian $H_s$ and jump operators $L_{e\rightarrow \downarrow}^{(k)}$ and $L_{e\rightarrow \uparrow}^{(k)}$ given in Eqs.~\eqref{eq:continuous_hamiltonian} and~\eqref{eq:continuous_jumps}. The color code and the letters make reference to the drives used in the discrete scenario (cf. Fig~\ref{fig:scheme}). (Right) The diagram represents the effective system obtained by first considering $\ket{ee}$ as the excited subspace and $\{\ket{\chi_{+}},\ket{\chi_{0}},\ket{\chi_{-}}\}$ as the dressed state of the triplet subspace. We then apply the effective operator formalism~\cite{Reiter2012} which introduces new couplings (purple arrows) and effective jump operators (orange arrows). The latter ones are expressed in Eq.~\eqref{eq:effective_jumps}. Colors and lengths of the arrows are uncorrelated with the strength of the couplings.}
    \label{fig:scheme_continuous}
\end{figure*}

To obtain the bit-flip probability $p$ for the phenomenological model in Fig.~\ref{fig:data}c), we perform the following measurements. First, we apply 16 cycles of the singlet preparation protocol. This ensures the motional state of the ions during the trial application of the drive (A) matches that of the last actual application of that drive during the sequence. Then, we reset the spin state to $\ket{\downarrow\downarrow}$ by optical pumping. Afterward, we apply a single round of the drive (A) and measure the probability to find exactly one ion in the $D_{5/2}$ state. This probability equals the bit-flip probability $p$. A second method we use is to, following optical pumping to $\ket{\downarrow\downarrow}$, apply an MS gate and ground state mapping pulses to prepare $\ket{\Psi^{+}}$. Applying a single round of drive (A) to $\ket{\Psi^{+}}$ excites one ion to $D_{5/2}$ with probability $p/2$. Whenever possible, we use both of these methods, check their consistency, and average the results to produce a final estimate of $p$. In general, applying the drive (A) on two ions prepared in $\ket{\uparrow\uparrow}$ does not lead to observable repumping. The only exception is Fig.~\ref{fig:errors} d), where a changing radial mode tilt causes an off-resonant excitation from $\ket{\uparrow}$ to the $D_{5/2}$ manifold. This means that $p/2$ underestimates the total excitation probability of the drive (A) out of $\ket{\Psi^{-}}$. To correct for that error, we measure separately the probability $p_{\downarrow\downarrow}$ of the drive (A) to cause a single-ion excitation out of $\ket{\downarrow\downarrow}$, and the probability $p_{\uparrow\uparrow}$ of a single-ion excitation of out $\ket{\uparrow\uparrow}$. The total bit-flip probability $p$ in Fig.~\ref{fig:errors} d) is then set to $p=p_{\downarrow\downarrow}+p_{\uparrow\uparrow}$.

\noindent
\section{Continuous protocol}
\label{sec:continuous_scenario}

As mentioned in the main text, this state preparation scheme has also a continuous counterpart. In this scenario, all the three drives (A), (B), and (C) are turned on simultaneously and for a specific choice of parameters the system will dynamically converge towards $\ket{\Psi^{-}}$. In this section, we explain this continuous framework and derive an empirical formula for the convergence rate of the protocol using the effective operator formalism~\cite{Reiter2012}.

Similarly to the sequential application of the drives, a continuous generation of the singlet state requires a collective excitation $S_{x,e}^2$. The interaction Hamiltonian (A) reduces to such a coupling in the weak-field regime $\delta\gg\eta\Omega$, i.e. when the population transfer to the intermediate levels with another vibrational quantum number is negligible. In this limit, the system's Hamiltonian is given by
\begin{align} \label{eq:continuous_hamiltonian}
    \begin{split}
        H_s &= \hbar\,J \, S_{x,e}^2 
            + \hbar\,\frac{\Omega_C}{2} \Big( \sigma_x \otimes \bm{1} + \bm{1} \otimes \sigma_x \Big) + \\
            &+\hbar\,\beta\,\Big(\ket{\uparrow}\bra{\uparrow} \otimes \bm{1} + \bm{1} \otimes \ket{\uparrow}\bra{\uparrow}\Big)
    \end{split}
\end{align}
where $J=\frac{(1/2\,\eta\Omega)^2}{\delta}$ represents the effective frequency of the collective excitation from $\ket{\downarrow \downarrow}$ to $\ket{ee}$; $\Omega_C$ the frequency of drive (C) and $\beta$ the detuning of the spin qubit. We model the repump (B) as a Markovian process with jump operators
\begin{equation}\label{eq:continuous_jumps}
    \begin{split} 
        L_{e\rightarrow \downarrow}^{(1)} &= \sqrt{p_{e\rightarrow \downarrow} \, \kappa}\, 
                                             \ket{\downarrow}\bra{e} \otimes \bm{1}\, , \\
        L_{e\rightarrow \uparrow}^{(1)} &= \sqrt{p_{e\rightarrow \uparrow} \, \kappa}\, 
                                             \ket{\uparrow}\bra{e} \otimes \bm{1} \, ,
    \end{split}    
\end{equation}
for the first ion and similarly for the second. Here, $p_{e\rightarrow \downarrow}$ ($p_{e\rightarrow \uparrow}$) represent the probability that an ion decays from $\ket{e}$ to $\ket{\downarrow}$ ($\ket{\uparrow}$) and can be, as in the discrete scenario, parametrized by $p_{e\rightarrow \downarrow}/p_{e\rightarrow \uparrow} = \tan^2(\gamma)$. We must, however, introduce an additional parameter $\kappa$ representing the repump rate. Finally, we assume that the evolution of the system is governed by the Lindblad master equation:
\begin{align} \label{eq:master_eq}
    \frac{\dd}{\dd t} \rho(t) = - \frac{i}{\hbar} [H_s,\rho(t)] + \sum_{j=1}^{4} \mathcal{D}[L_j](\rho(t)) \, ,
\end{align}
where
\begin{align} \label{eq:dissipator}
    \mathcal{D}[L_j](\rho(t)) = L_j\,\rho(t)\,L_j^\dagger - 
    \frac{1}{2} \left\{ L_k^\dagger\,L_k, \rho(t) \right\} \, ,
\end{align}
is the dissipator for a given jump operator $L_j$. The energy levels of the system in the two-ion basis are depicted in Fig.~\ref{fig:scheme_continuous}. On top of the usual collective spin, $\{ \ket{\downarrow\downarrow},\,\ket{\uparrow \uparrow},\,\ket{\Psi^+},\,\ket{\Psi^-}\}$, and excited, $\ket{ee}$, states, the figure also shows the intermediate, single-excitation states which, in this continuous framework, play a non-negligible role. The objective is now to determine the right balance of the parameters $(J,\Omega_C,\gamma,\kappa)$ which would lead to the desired steady state. The quantity of interest is then the fidelity of $\ket{\Psi^-}$ given by 
\begin{align} \label{eq:fidelity_continuous}
    F(t) \approx 1 - C_0 \, e^{- R_t \, t} \,
\end{align}
with $R_t$ being the convergence rate in unit of time and ultimately representing the Liouvillian gap of the system given in Eq.~\eqref{eq:master_eq}.

To begin with, we can argue that the strength of the drives (B) and (C) should be larger or at least comparable to $J$, since in an optimally converging protocol immediately after the excited state $\ket{ee}$ gets populated, it must be directly repumped to lower-excitation states. Similarly for the common rotation $\Omega_C$, the ground “spin” state manifold is expected to be shuffled more frequently than collectively excited with $S_{x,e}^2$ such that $\ket{\downarrow\downarrow}$ is never depleted. Finally, the additional freedom given by the detuning $\beta$ allows us to compensate for the AC Stark shift (and thus having all the drives on resonance) by setting it to $\beta=J$. This value of the detuning has been numerically verified to optimize $R_t$ and has been thus kept constant in our following derivations.

With that in mind, we can identify $\ket{ee}$ as the excited subspace and thus apply the effective operator formalism to reduce the evolution to the ground-state dynamics only~\cite{Reiter2012}. In order to remain as general as possible, we move to the dressed basis of the drive (C), 
\begin{equation}\label{eq:dressed_states}
    \begin{split} 
        \ket{\chi_{0}} &= \frac{1}{\sqrt{2}} \Big(\ket{\downarrow\downarrow} - \ket{\uparrow\uparrow}\Big) \,\, , \\
        \ket{\chi_{\pm}} &= \frac{1}{2} \Big(\ket{\downarrow\downarrow} + \ket{\uparrow\uparrow} \pm \sqrt{2} \ket{\Psi_{+}} \Big)\,\,. \\
    \end{split} 
\end{equation}

Fig.~\ref{fig:scheme_continuous} illustrates the energy levels and couplings of the new effective system. With the absence of the excited level, the dressed states have now some modified couplings (purple arrows)
\begin{equation}\label{eq:effective_hamiltonian}
    \scalebox{.93}{$
    \begin{split} 
        H_\mathrm{\chi}^{\mathrm{eff}}\!=& \,
        \hbar\,\Omega_C\left(1 + \frac{J^2}{\kappa^2+\Omega_C^2}\right)\!
        \big(\!\ket{\chi_{+}}\bra{\chi_{+}} - \ket{\chi_{-}}\bra{\chi_{-}}\!\big)+ \\
        +&\,\hbar\,\frac{\Omega_C}{2}\,\frac{\sqrt{2}\,J^2}{\kappa^2-i\kappa\Omega_C}\,\ket{\chi_0}\bra{\chi_{+}} \,+\, \mathrm{H.c.} \,+ \\
        -&\,\hbar\,\frac{\Omega_C}{2}\,\frac{\sqrt{2}\,J^2}{\kappa^2+i\kappa\Omega_C}\,\ket{\chi_0}\bra{\chi_{-}} \,+\, \mathrm{H.c.}
    \end{split} 
    $}
\end{equation}
and decay directly into the single-excited states with effective jump operators (orange arrows)
\begin{equation}\label{eq:effective_jumps}
    \scalebox{.93}{$
    \begin{split} 
        L_{1}^{\mathrm{eff}} &= \sqrt{\cos^2(\gamma) \kappa} \ket{\uparrow e}
                                \big(C_{+} \bra{\chi_{+}} + C_{0} \bra{\chi_{0}} + C_{-} \bra{\chi_{-}}\!\big) \\
        L_{2}^{\mathrm{eff}} &= \sqrt{\cos^2(\gamma) \kappa} \ket{e \uparrow}
                                \big(C_{+} \bra{\chi_{+}} + C_{0} \bra{\chi_{0}} + C_{-} \bra{\chi_{-}}\!\big) \\
        L_{3}^{\mathrm{eff}} &= \sqrt{\sin^2(\gamma) \kappa} \ket{\downarrow e}
                                \big(C_{+} \bra{\chi_{+}} + C_{0} \bra{\chi_{0}} + C_{-} \bra{\chi_{-}}\!\big) \\
        L_{4}^{\mathrm{eff}} &= \sqrt{\sin^2(\gamma) \kappa} \ket{e \downarrow}
                                \big(C_{+} \bra{\chi_{+}} + C_{0} \bra{\chi_{0}} + C_{-} \bra{\chi_{-}}\!\big)
    \end{split} 
    $}
\end{equation}
with constants
\begin{equation}\label{eq:effective_jumps_constants}
    C_{0} = \frac{ \sqrt{2}\, J }{- i \kappa} \qquad
    C_{\pm} = \frac{J}{\mp\,\Omega_C - i \kappa } \, .
\end{equation}
The reader can notice that the effective evolution introduces some modified detunings and some new effective Rabi frequencies between $\ket{\chi_0}$ and $\ket{\chi_{\pm}}$ which in the original collective spin basis results in couplings between all the triplet states.

\begin{figure}[t]
    \centering
    \includegraphics[trim=5pt 5pt 5pt 5pt, clip]{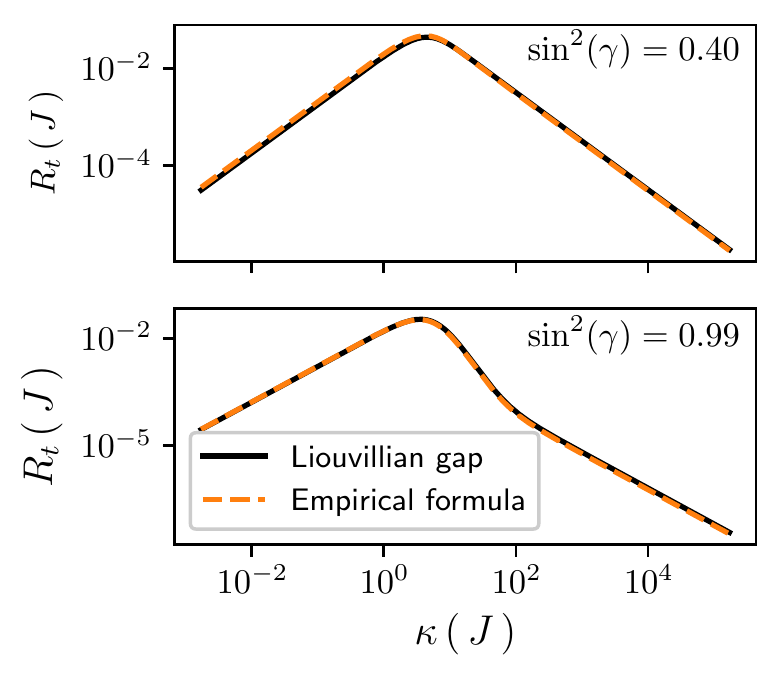}
    \caption{Protocol's convergence rate $R_t$ as a function of the repump rate for two different branching ratio $\gamma$. The empirical curve is calculated using Eq.~\eqref{eq:conv_rate_continuous} (with a proportionality factor of $2.4$) and the Liouvillian gap was obtained numerically from the Lindblad master equation~\eqref{eq:master_eq}. The frequency of the drive (C) as been kept constant and equal to $\Omega_C=6J$.}
    \label{fig:kappa_Rt}
\end{figure}

Even though we eliminated one state out of nine, the problem remains convoluted and the eigenanalysis of the effective Liouvillian is analytically difficult to perform. However, an empirical approach can help us to estimate the convergence rate of the protocol. Numerical observations of the behavior of the Liouvillian gap for different values of $(\Omega_C,\kappa,\gamma)$ allows us to derive the following expression
\begin{equation} \label{eq:conv_rate_continuous}
    \begin{split} 
        R_t \propto &\,\Big(\!\cos^4(\gamma) + \sin^4(\gamma)\Big)\,\kappa\,
                        \left(\frac{\Omega_C}{2}\frac{J}{\kappa^2 + \Omega_C^2}\right)^2 \\
                    &+ \sin^2(\gamma)\cos^2(\gamma)\,\kappa\,\frac{J^2}{\kappa^2 + \Omega_C^2}.
    \end{split}
\end{equation}

This formula has been obtained by considering the slowest process given by the decay channels $\ket{\chi_{-}}\rightarrow\ket{\Psi_{-}}$ and $\ket{\chi_{+}}\rightarrow\ket{\Psi_{-}}$ and distinguishing two regimes of the system's dynamics: with a highly uneven branching ratio, $|\gamma-\pi/4|\approx\pi/4$, and with a comparable decay, $|\gamma-\pi/4|\approx 0$. In the first limit, the preparation of the singlet state is slow, because the dynamics is dominated by ``second-order processes". This means that the system must undergo a collective rotation before decaying into the singlet state. This corresponds to the first term in the parenthesis of Eq.~\eqref{eq:conv_rate_continuous}. The second limit coincides with a fast, first-order preparation of $\ket{\Psi_{-}}$. It is given by the second term of $R_t$ and can be rewritten as $p_{e\rightarrow \downarrow}\,p_{e\rightarrow \uparrow}$ times the decay rate $\kappa\,|C_{-}|^2$. The probabilities intuitively represent the process $\ket{ee}\rightarrow\ket{\Psi^{-}}$: one of the excited ions should decay into a down state and the other into an up state. This regime is also valid for highly uneven branching ratio when $\kappa$ becomes comparable to $\sin^2(\gamma)$. Finally, the proportionality coefficient for $R_t$ is equal to $2.4$ and has been determined numerically using the simulated Liouvillian gap. Fig.~\ref{fig:kappa_Rt} shows the comparison between the empirical convergence rate given by Eq.~\eqref{eq:conv_rate_continuous} and the numerical value of the Liouvillian gap for different repump rates $\kappa$ and branching ratios $\gamma$.

We must mention that the empirical formula presented above is valid as long as the sum of the frequency of the drive (C) and the repump rate is larger than the frequency of the collective excitation $J$. Conversely, the population in the state $\ket{ee}$ could not be discarded anymore using the effective operator formalism.

\begin{figure}[t]
    \includegraphics[trim=5pt 5pt 5pt 5pt, clip]{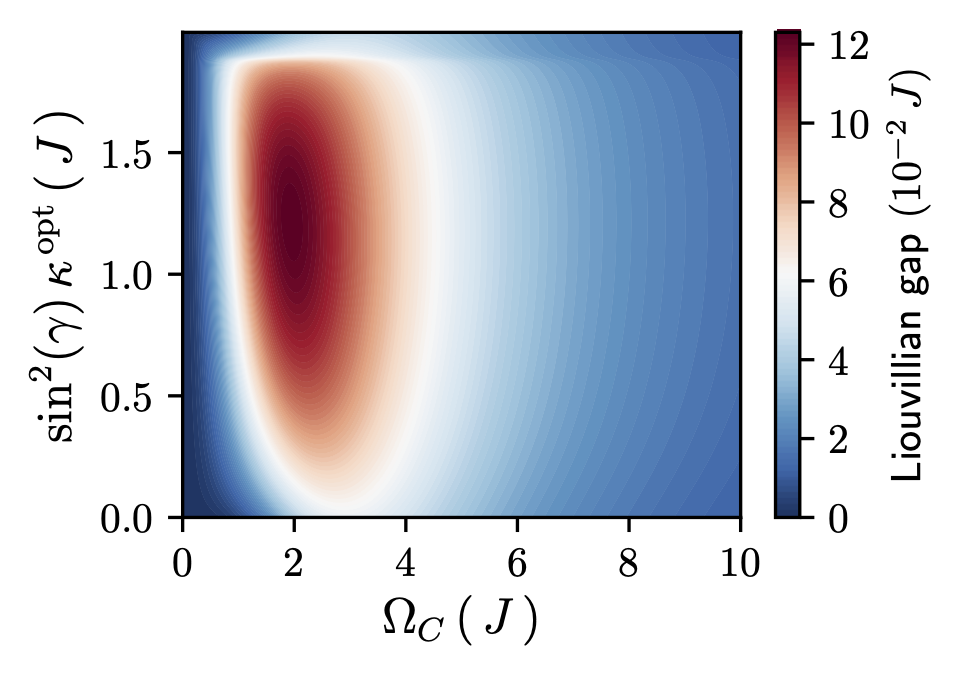}
    \caption{Liouvillian gap of the system given in Eq.~\eqref{eq:master_eq} for different frequencies $\Omega_C$ of drive (C) and repump rates. To reduce the dimensionality of the parameter space of the continuous scenario, we first optimized the Liouvillian gap for various $p_{e\rightarrow \downarrow}=\sin^2(\gamma)\in(0,1)$ with respect to $(\Omega_C,\kappa)$. The vertical axis of the figure has then been obtained by the product of $p_{e\rightarrow \downarrow}$ with the corresponding optimal repump rate $\kappa^\mathrm{opt}$.}
    \label{fig:liouvillian_gap}
\end{figure}

We are now interested in the optimal value of this convergence rate and the parameters which lead to it. Fig.~\ref{fig:liouvillian_gap} shows the Liouvillian gap as a function of $\Omega_C$ and $\sin^2(\gamma)\,\kappa^\mathrm{opt}$ where the optimal repump rate has been obtained from a multivariate minimization with respect to $(\Omega_C, \gamma)$ for different branching ratio such that $0<\sin^2(\gamma)<1$. We thus obtain that $R_t^\mathrm{opt}=6.1\times10^{-2}\,J$ when $\Omega_C$, $\kappa$ and $\gamma$ are equal to $1.95\,J$, $2.58\,J$ and $0.29\pi$ respectively. For a collective excitation of frequency $J=0.58\,\mathrm{kHz}$ (i.e. $\delta=20\eta\Omega$), this would correspond to a convergence rate of $72\,\mathrm{Hz}$.


}{}

\end{document}